\documentclass[12pt]{article}
\usepackage{latexsym,epsfig,graphicx,a4wide,amssymb,cite}

\def\a{\alpha}
\def\b{\beta}
\def\g{\gamma}

\def\d{\delta}

{\rm }

\def\La{\Lambda}

\def\m{\mu}
\def\n{\nu}
\def\r{\rho}

\def\vep{\varepsilon}

\def\th{\theta}

\def\bR{\bar{R}}
\def\bT{\bar{T}}
\def\bg{\bar{g}}
\def\bH{\bar{H}}
\def\be{\begin{equation}}
\def\ee{\end{equation}}
\def\bea{\begin{eqnarray}}
\def\eea{\end{eqnarray}}
\def\nn{\nonumber}
\newcommand{\lb}{\label}

\def\2{\frac{1}{2}}
\def\4{\frac{1}{4}}

\catcode`\@=11

\@addtoreset{equation}{section}

\def\@normalsize{\@setsize\normalsize{15pt}\xiipt\@xiipt
\abovedisplayskip 14pt plus3pt minus3pt%
\belowdisplayskip \abovedisplayskip
\abovedisplayshortskip  \z@ plus3pt%
\belowdisplayshortskip  7pt plus3.5pt minus0pt}
\def\small{\@setsize\small{13.6pt}\xipt\@xipt
\abovedisplayskip 13pt plus3pt minus3pt%
\belowdisplayskip \abovedisplayskip
\abovedisplayshortskip  \z@ plus3pt%
\belowdisplayshortskip  7pt plus3.5pt minus0pt
\def\@listi{\parsep 4.5pt plus 2pt minus 1pt
            \itemsep \parsep
            \topsep 9pt plus 3pt minus 3pt}}

\def\underline#1{\relax\ifmmode\@@underline#1\else
        $\@@underline{\hbox{#1}}$\relax\fi}
\@twosidetrue \relax

\catcode`@=12

\catcode`\@=11

\def\section{\@startsection{section}{1}{\z@}{3.5ex plus 1ex minus
   .2ex}{2.3ex plus .2ex}{\large\bf}}
\def\ps@headings{\def\@oddfoot{}\def\@evenfoot{}
\def\@oddhead{\hbox{}\hfill
        \makebox[.5\textwidth]{\raggedright\ignorespaces --\thepage{}--
        \hfill }}
\def\@evenhead{\@oddhead}
\def\subsectionmark##1{\markboth{##1}{}}
}

\ps@headings

\catcode`\@=12

\begin{document}

\begin{titlepage}

\rightline{December 2010}

\begin{centering}
\vspace{1cm}
%i
{\Large {\bf Perturbations of  Gauss-Bonnet Black Strings in Codimension-2 Braneworlds}}\\

\vspace{1.5cm}

 {\bf Bertha Cuadros-Melgar $^{*}$},
 {\bf Eleftherios Papantonopoulos}$^{**}$, {\bf Minas~Tsoukalas} $^{\flat}$ \\
 \vspace{.2in}
 Department of Physics, National Technical University of
Athens, \\
Zografou Campus GR 157 73, Athens, Greece. \\
 \vspace{.2in}

  {\bf Vassilios Zamarias}$^{\natural}$\\
 \vspace{.2in}
Hellenic Army Academy, Vari GR 166 73, Attica, Greece, \\
and Department of Physics, National Technical University of
Athens, \\
Zografou Campus GR 157 73, Athens, Greece. \\

\vspace{3mm}

\end{centering}
\vspace{2cm}

\begin{abstract}

We derive the  Lichnerowicz equation in the presence of the
Gauss-Bonnet term. Using the modified Lichnerowicz equation we
study the metric perturbations of  Gauss-Bonnet black strings in
Codimension-2 Braneworlds.

\end{abstract}

\vspace{2.5cm}
\begin{flushleft}
$^{*}~~$ e-mail address: berthaki@gmail.com \\
$^{**} ~$ e-mail address: lpapa@central.ntua.gr \\
$ ^{\flat}~~$ e-mail address: minasts@central.ntua.gr\\
$ ^{\natural}~~$ e-mail address: zamarias@central.ntua.gr

\end{flushleft}
\end{titlepage}

\section{Introduction}

Higher-dimensional black holes and black strings appear as
classical solutions of gravity theories in more than four
dimensions. One central issue concerning their behaviour is the
stability issue. Recently, the stability of higher-dimensional
black holes and in particular of black strings, is under intense
investigation. The stability of higher-dimensional black holes has
been addressed adequately in
 \cite{Kodama:2003jz}. Also, the understanding of the
stability of black rings and in general of higher-dimensional
black objects has also been improved considerably (for a review
see \cite{Harmark:2007md,Obers:2008pj}).

In the case of black holes that  are solutions of gravity theories
with high order curvature terms \cite{chaboul,Wheeler:1985qd,
Wiltshire:1988uq,chawiltshire0, Crisostomo:2000bb,Aros:2000ij}, the stability issue
is an intriguing one. The Lovelock theory \cite{Lovelock:1971yv}
is a non-trivial extension of General Relativity which apart from
the Einstein-Hilbert term also  includes higher order curvature
terms. In fact, the Lovelock theory contains terms only up to the
second order derivatives in the equations of motion which makes
the theory more tractable. In spite of that, the black hole
solutions of Lovelock theory could be found mostly because they
are highly symmetric (for a review see \cite{Charmousis:2008kc},\cite{Garraffo:2008hu}).
 The study of their stability requires the application of
linear perturbation theory, which however, confronts  with the
high complexity of the Lovelock equations, since the perturbative
terms break the simplifying symmetries of the background metric.

In the simplest case of a Gauss-Bonnet theory which is a second
order Lovelock theory, the classical stability of black hole
solutions has been studied. The stability analysis under  scalar,
 vector and tensor perturbations has been
performed~\cite{Dotti:2004sh,Dotti:2005sq,Gleiser:2005ra}. It was
found that there exists a scalar mode instability in five
dimensions, a tensor mode instability in six dimensions, and no
instability in other dimensions. Recently the master equations of
Lovelock black holes of scalar, vector and tensor perturbations
were derived in
\cite{Takahashi:2009dz,Takahashi:2009xh,Takahashi:2010ye} and it
was shown  that small Lovelock black holes are unstable
 in the asymptotically flat cases~\cite{Takahashi:2010gz}.

 Higher order
curvature terms are  known to appear in string theory, generically
introducing higher derivatives in the metric~\cite{string1}. This
will inevitably lead to perturbative ghosts. However, the second
order curvature terms are known to take the form of the
Gauss-Bonnet combination~\cite{gross}. This corresponds to the
non-trivial second order Lovelock theory which has no higher
derivatives in the effective string action, so we expect no ghosts
to appear in second order. However, it was found that spherically
symmetric vacuum solutions in the Gauss-Bonnet theory   suffer
from ghost-like instabilities and it was conjectured that these
instabilities persist in all vacuum Lovelock
solutions~\cite{Charmousis:2008ce}. In fact it was shown that
there is a limit, the Chern-Simons limit, in which the theory
becomes strongly coupled and linear perturbation theory breaks
down. In this limit it was also shown, that if there is a
fine-tuning between the parameters, the two branches of vacuum
solutions  coincide with the Chern-Simons black hole solution
which has maximum symmetry. Scalar perturbations of the
Chern-Simons black holes were studied in \cite{Gonzalez:2010vv}.

Higher order curvature terms  also appear in braneworlds and
especially in codimension-2 braneworld models which were mainly
introduced because there is a mechanism of self-tuning of the
effective cosmological constant  to zero \cite{6d}. However, soon
it was realized \cite{Cline}
   that one can only find nonsingular
solutions if the brane energy momentum tensor is proportional to
its induced metric. To reproduce the effective four-dimensional
Einstein equation  on the brane one has  to modify the
gravitational action  by the inclusion of a Gauss-Bonnet
term~\cite{Bostock:2003cv,Papantonopoulos:2005ma} (for a review
see \cite{Papantonopoulos:2006uj}) or to regularize  the conical
singularities (see \cite{Papantonopoulos:2006dv} and references
therein).

Recently   black hole solutions of codimension-2 braneworlds have
been found. A six-dimensional black hole localized on a 3-brane of
codimension-2 was proposed in \cite{Kaloper:2006ek}.  However, it
is not clear how to realize these solutions in the thin brane
limit where high order curvature terms are needed to accommodate
matter on the brane. Generalizations to include rotations were
presented in~\cite{Kiley:2007wb} and perturbative analysis of
these black holes were carried out in~\cite{Kaloper:2007ap,
alBinni:2007gk,Chen:2007ay}.

In the case that there is a Gauss-Bonnet term in the bulk, black
hole solutions were studied in a five-dimensional codimension-2
braneworld model~\cite{CuadrosMelgar:2007jx}. Two classes of
solutions were found. The first class consists of the familiar BTZ
black hole \cite{Banados:1992wn} which solves the junction
conditions on a conical 2-brane in vacuum. These solutions in the
bulk are BTZ string-like objects with regular horizons and no
pathologies.  The second class of solutions consists of BTZ black
holes with short distance corrections. These solutions correspond
to a BTZ black hole conformally dressed with a scalar
field~\cite{Zanelli1996,Henneaux:2002wm} and they  have black
string-like extensions into the bulk. Generalizations to include
angular momentum and charge were presented in
\cite{CuadrosMelgar:2009qb}. Also four-dimensional
Schwarzschild-AdS black hole solutions on the brane were found,
which in the six-dimensional spacetime look like black string-like
objects with regular horizons \cite{CuadrosMelgar:2008kn}. The
warping to extra dimensions depends on the Gauss-Bonnet coupling
which is fine-tuned to the six-dimensional cosmological constant.
The presence of the Gauss-Bonnet term in codimension-2 braneworlds
has important consequences. Its projection on the brane gives a
consistency relation~\cite{Papantonopoulos:2005ma} that dictates
the form of the solutions. It allows black string solutions in
five dimensions, and in six dimensions it specifies the kind of
matter which is needed  in the bulk in order to support a black
hole solution on the brane. We note here that black string
solutions with a Gauss-Bonnet term in the bulk are not possible in
codimension-1 braneworlds \cite{Barcelo:2002wz}.

The stability analysis of these static black hole solutions of
codimension-2 braneworlds is interesting. They solve N-dimensional
Einstein field equations with the presence of a Gauss-Bonnet term
in the action. The symmetries of the solutions are
$\mathcal{M}^{d} \times \mathcal{K}^{n-d}$ \footnote{Black hole
solutions with a Gauss-Bonnet term and with these kind of
symmetries were discussed in \cite{Molina:2008kh} }, where
$\mathcal{M}$ is a maximally symmetric space while the space
$\mathcal{K}$ is axially symmetric. Therefore, any stability
analysis of these static solutions, unlike the Lovelock black hole
solutions which in general are spherically symmetric, has also to
confront with the particular symmetries of the solutions.

In this work we address the problem of stability of codimension-2
black strings with a Gauss-Bonnet term in the bulk and more
generally of gravity theories with high order curvature terms that
have time-independent solutions with symmetries other than
spherical.\footnote{The stability of uniform black string
solutions in the Gauss-Bonnet theory was studied in
\cite{Brihaye}.} In codimension-1 braneworlds the Schwarzschild
metric on the brane was considered and  its black string extension
into the bulk~\cite{Chamblin:1999by} was studied. It was found
that  this string is unstable to classical linear
perturbations~\cite{BSINS}. A lower dimensional version of a black
hole living on a (2+1)-dimensional braneworld was considered
in~\cite{EHM}. A BTZ black hole on the brane was found which can
be extended as a BTZ black string in a four-dimensional AdS bulk.
Their thermodynamical stability analysis showed that the black
string remains a stable configuration when its transverse size is
comparable to the four-dimensional AdS radius, being destabilized
by the Gregory-Laflamme instability~\cite{BSINS} above that scale,
breaking up to a BTZ black hole on a 2-brane. The stability of BTZ
black string  was also discussed in \cite{Bin}.

One way to study linear stability of higher dimensional objects
with curvature corrections is to find the eigenvalues of the
Lichnerowicz operator of a given perturbation
\cite{Gibbons:2002pq}. This method has the advantage of being
formulated in a gauge invariant way allowing the study of metrics
with various symmetries. We will derive the Lichnerowicz equation
in the presence of the Gauss-Bonnet term. We will show that a
simple application of the modified Lichnerowicz equation to the
case of D=6 spherically symmetric black hole solutions of
Gauss-Bonnet theory can easily reproduce the known results
of~\cite{Dotti:2004sh} for tensor perturbations for these black
holes.

We will subsequently use the modified Lichnerowicz equation to the
study of linear perturbations of five-dimensional black string
solution of codimension-2 braneworld model of
\cite{CuadrosMelgar:2007jx}, away from the Chern-Simons limit. We
will show that the knowledge of the Lichnerowicz equation can
provide very important information on the stability analysis  of a
complex  system of coupled differential equations that has to be
solved. In the case of codimension-2 geometries, the black string
can propagate in two transverse extra dimensions so intuitively,
one expects that metric perturbations to have more severe
stability problems than the conventional five-dimensional black
strings propagating in one extra transverse dimension.

Our analysis shows that for tensor perturbations the modified
Lichnerowicz equation, due to the  symmetries of the codimension-2
black string solution, can not give any information on the
perturbed system, indicating that either there is no tensor
propagating modes or we are in a strong coupling regime.  For the
vector perturbations we find a degeneracy of the modes on the bulk
space having no dependence on the brane coordinates indicating an
instability on the modes generated on the brane and propagating
into the bulk. To calculate the scalar perturbations, we will
apply the derived modified Lichnerowicz equation to the scalar
part of the metric perturbations. We will show that, due to the
fact that the metric of the black string solutions can be brought
to  a factorizable form, the results are the same as having
studied the propagation of a scalar field in the background metric
of the black string, by solving the Klein-Gordon equation. We find
stability for the scalar modes by both methods.

Another important information that the modified Lichnerowicz
equation can give us is the behaviour of the theory in the
Chern-Simons limit. As we have already discussed, in the
Gauss-Bonnet theory there are two branches of spherically
symmetric vacuum solutions: The Schwarzschild-AdS branch (known as
the Einstein limit) and the string branch (known as the
Gauss-Bonnet limit)~\cite{Charmousis:2008ce}. Both branches
coincide at the Chern-Simons limit where the theory becomes
strongly coupled. As we will discuss in the following, in our case
the Chern-Simons limit manifest itself as a prefactor in front of
the modified Lichnerowicz equation. The limit where this prefactor
goes to zero is an indication of a strong coupling problem,
signaling that linear perturbation theory breaks down.

 The paper is organized as
follows. In section 2 we  present the general formalism for linear
perturbations and derive the Lichnerowicz equation. We further
generalize this formalism by including the Gauss-Bonnet term  and
derive the modified Lichnerowicz equation where  a source term
appears, due to the presence of the Gauss-Bonnet term. In section
3 we apply these results to spherically symmetric solutions of the
Gauss-Bonnet theory. In section 4 we calculate the scalar
perturbations of the five-dimensional black hole solution of the
codimension-2 braneworld model and we discuss its  vector and
tensor perturbations. Finally, in section 5 we conclude.
%\newpage
%%%%%%%%%%%%%%%%%%%%%%%%%%%%%%%%%%%%%%%%%%%%%%%%%%%%%%%%%%%%%%%%%%%%%%%%

\section{The Lichnerowicz Equation}

In this section we will review the general formalism of linear
metric perturbations and we will derive  the D-dimensional
Lichnerowicz equation \cite{Gibbons:2002pq} and the D-dimensional
modified Lichnerowicz equation in the presence of a Gauss-Bonnet
term.

\subsection{General formalism for linear metric perturbations}

Consider a $D$-dimensional Einstein manifold
$(\mathcal{B},\tilde{g}) $. The stability analysis can be reduced
in finding  a solution of an ordinary differential equation of a
Schr\"{o}dinger type, {\it i.e.}, the Lichnerowicz equation. This
amounts to determine the spectrum of the Lichnerowicz operator on
transverse traceless symmetric tensor fields of the manifold
$\mathcal{B}$. We will first expose the general formalism and then
we will apply it to a manifold with a Gauss-Bonnet term.

We start from the  definition  for the Riemann Tensor as in
\cite{Wald:1984rg},
\be
\nabla_{C} \nabla_{D} T^{A}-\nabla_{D}
\nabla_{C} T^{A}=R^{A}_{BCD}T^{B}
\ee
and consider the following
linear perturbation of a $D$-dimensional metric background
$\bar{g}_{MN}$
\be
g_{MN}=\bg_{MN} + \vep\,h_{MN},
\ee
where the
Latin capital indices $M$,~$N$ take values in the $D$-dimensional
space, and all unperturbed quantities are written as $\bar{X}$. We
also decompose the Ricci tensor and a $D$-dimensional energy
momentum tensor $T_{MN}$  respectively as
\bea
R_{MN}&=&\bR_{MN}+\vep\,\d R_{MN}, \\
T_{MN}&=&\bT_{MN}+\vep\,\d T_{MN}. \eea The Einstein's field
equations are \be G_{MN}=R_{MN}-\2\,R\,g_{MN}=T_{MN},
\lb{Einstein1} \ee and using the trace they can  be written as \be
\lb{Einstein2} R_{MN}=T_{MN}-\2\,\frac{2}{D-2}\,g_{MN}\,T^L_L. \ee
The zeroth order Einstein equations (or background solution) are
\be \lb{0Einstein}
\bR_{MN}=\bT_{MN}-\2\,\frac{2}{D-2}\,\bg_{MN}\,\bT^L_L, \ee while
the first order Einstein equations (or perturbed equations) are \be
\lb{1Einstein} \d R_{MN}=\d S_{MN}, \ee where \be \lb{PRicci1} \d
R_{MN}=-\2\left(\Box h_{MN} + \nabla_N\nabla_M\,h -
\nabla^K\nabla_M\,h_{KN} - \nabla^K\nabla_N\,h_{MK}\right), \ee
with $\Box h_{MN} = \bg^{AB}\,\nabla_A\,\nabla_B h_{MN}$, and \be
\d S_{MN}=\d T_{MN} -
\2\,\frac{2}{D-2}\left(\bg_{MN}\,\bg^{KL}\,\d T_{KL} +
\bg_{MN}\,h^{PS}\,\bT_{PS} -h_{MN}\,\bT^L_L\right).
\ee
It is easy
to check that  the Einstein equations (\ref{Einstein1}) are
satisfied for $g_{MN}=\bg_{MN}+\vep\,h_{MN}$ and
$T_{MN}=\bT_{MN}+\vep\,\d T_{MN}$. To simplify the equations we
choose the de Donder gauge, namely, the traceless and the
transverse gauge conditions, respectively given by
\bea
\lb{TraceGauge}
\bg^{MN}\,h_{MN}=0,\\
\lb{TransverseGauge} \nabla^M\,h_{MN}=0.
\eea
Then, in this gauge
the first order Ricci tensor (\ref{PRicci1}) can be written in the
following form
\be \lb{PRicci} \d\,R_{MN}=-\2 \left( \Box\,h_{MN}
- 2\,\bR_{KNML}\,h^{KL} - \bR^K_M\,h_{KN} -
\bR^K_N\,h_{MK}\right).
\ee

In vacuum $(T_{MN}=0)$, the Einstein equation reduces to $R_{MN}=0$,
whose zeroth order part, $\bR_{MN}=0$, can be used in the first
order part (\ref{PRicci}) to obtain the \emph{Lichnerowicz equation
for linear perturbations in vacuum}
\be \Box\, h_{MN} -
2\,R_{KMNL}\,h^{KL} = 0. \lb{Lichne}
\ee

When a cosmological constant is present, {\it i.e.}, $(T_{MN}=-\La_D\,
g_{MN})$, the zeroth and first order parts of the Einstein
equations (\ref{Einstein2}) are respectively
\bea
\bR_{MN}=-\tilde{\La}_D\,\bg_{MN},
\lb{0EinsteinCosmo}\\
\d R_{MN}=-\tilde{\La}_D\,h_{MN}, \lb{1EinsteinCosmo}
\eea
where $\tilde{\La}_D = \frac{D-3}{D-2}\,\La_D$. Using
(\ref{0EinsteinCosmo}) in (\ref{PRicci}), the first order part
(\ref{1EinsteinCosmo}) gives again the Lichnerowicz equation
(\ref{Lichne}).

\subsection{The modified Lichnerowicz equation}

We want to establish the equivalent to the Lichnerowicz equation
in the case where a Gauss-Bonnet term is included in the manifold
$\mathcal{B}$. In five or in six  dimensions the Gauss-Bonnet
density term in the action is given by
\be
\a \mathcal{L}_{GB}=\a
\left(R^2-4R_{KL}\,R^{KL}+R_{KLPQ}R^{KLPQ}\right),
\ee
where $\a(\geq 0)$ is the Gauss-Bonnet coupling constant. Its variation
includes the Gauss-Bonnet term $\a\,H_{MN}$ in the field equations
with $H_{MN}$ being the Gauss-Bonnet tensor,

\be
G_{MN}- \a H_{MN}=-\La_D\,g_{MN}. \label{gbeq}
\ee

For convenience, we split this tensor in five terms
$H_{MN_i} (i=1,...,5)$ which are
\bea
H_{MN_1} &=& \2\,g_{MN}\,\mathcal{L}_{GB}, \\
H_{MN_2} &=& - 2R\,R_{MN}, \\
H_{MN_3} &=& + 4 R_{MK}\,R_N^K, \\
H_{MN_4} &=& + 4R^K_{MLN}\,R_K^L, \\
H_{MN_5} &=& - 2R_{MKLP}\,R_N^{KLP}.
\eea
As for the linear
perturbations, we decompose the Gauss-Bonnet  tensor as
\be
H_{MN}= \bH_{MN} + \vep\,\d H_{MN} = \sum_{i=1}^5 \left(\bH_{MN_i} +
\vep\,\d H_{MN_i}\right),
\ee
and after some lengthy calculations,
where we used the gauge conditions (\ref{TraceGauge}) and
(\ref{TransverseGauge}), we find that the first order Gauss-Bonnet
 tensor decompositions are
\bea
\lb{PH1} \d H_{MN_1}
&=& \2 \,h_{MN} \mathcal{L}_{GB} - \bg_{MN}\,h^{KL}\,\bR_{KL}\,\bR
+2\bg_{MN}\,\bR^{KL}\,\Box\,h_{KL} -
4\bg_{MN}\,\bR^{KL}\,\bR^P_{\,\,\,KL}\hspace{-0.04in}\,^Q\,h_{QP}
\nn \\ &~&-\bg_{MN}\,\bR^{KLPQ}\,h_{KR}\,\bR^R_{\,\,\,LPQ}
+ 2\bg_{MN}\,\bR^{KLQP}\,\nabla_Q\nabla_L\,h_{KP}, \\
\lb{PH2}
\d H_{MN_2} &=& 2h^{KL}\,\bR_{KL}\,\bR_{MN} + \bR\,\Box\,h_{MN} - 2\bR\,\bR^K_{\,\,\,MN}\hspace{-0.04in}\,^L\,h_{LK} \nn \\
&~&-\bR\,\bR^K_{\,\,\,N}\,h_{MK} - \bR\,\bR^K_{\,\,\,M}\,h_{NK}, \\
\lb{PH3}
\d H_{MN_3} &=& 2\bR^K_{\,\,\,N}\left(-\Box\,h_{MK} + 2\bR^L_{\,\,\,MK}\hspace{-0.04in}\,^Q\,h_{LQ} + \bR^L_{\,\,\,K}\,h_{ML}\right) \nn \\
&~&+2\bR^K_{\,\,\,M}\left(-\Box\,h_{KN} + 2\bR^L_{\,\,\,NK}\hspace{-0.04in}\,^Q\,h_{LQ} + \bR^L_K\,h_{LN}\right), \\
\lb{PH4}
\d H_{MN_4} &=& 2\bR^{KL} \left(-\bR^P_{\,\,\,KLN}\,h_{MP} + \nabla_L\nabla_M\,h_{KN} - \nabla_L\nabla_K\,h_{MN} - \nabla_N\nabla_M\,h_{KL} + \nabla_N\nabla_K\,h_{ML} \right) \nn \\
&~& + 2\bR_{KMLN} \left(-h^{LP}\,\bR^K_{\,\,\,P} - \Box\,h^{KL} - 2\bR^{LPKQ}\,h_{QP} \right),\\
\lb{PH5}
\d H_{MN_5} &=& +2\bR_N^{\,\,\,KLP} \left(\nabla_P\nabla_K\,h_{ML} - \nabla_P\nabla_M\,h_{KL} \right) + 2\bR_M^{\,\,\,KLP} \left(\nabla_P\nabla_K\,h_{LN} - \nabla_P\nabla_N\,h_{KL} \right) \nn \\
&~&-\bR_{MKLP}\,\bR^{QKLP}\,h_{QN} +
4\bR_M^{\,\,\,KLP}\,\bR_{NKQP}\,h^Q_{\,\,\,L} -
\bR_N^{\,\,\,KLP}\,\bR_{QKLP}\,h^Q_{\,\,\,M}.
 \eea
%while the trace
%of the first order Gauss-Bonnet  tensor is \bea
%\d H &=& -3\bR\,\bR^{KL}\,h_{KL} + 6 \bR^{KL}\,\Box\,h_{KL} - 16 \bR^{KLPQ}\,\bR_{KP}\,h_{LQ} -6 \bR^{KLPQ}\nabla_Q\nabla_L\,h_{KP} \nn \\
%&~& + 3 \bR^K_{\,\,\,LPQ}\,\bR^{RLPQ}\,h_{KR} + 4
%\bR^P_K\,\bR^Q_P\,h^K_Q - 2 \bR_{KLPQ}\,\bR^{RLPQ}\,h^K_R.
%\lb{TracePH} \eea
The Einstein's field equations (\ref{gbeq}) can also be written as
\be R_{MN}=\a\,H_{MN} -\La_D\,g_{MN} -
\frac{1}{D-2}\,g_{MN}\,\left(\a\,H - D\,\La_D \right), \ee where
the zeroth order equations are \be \lb{0EinsteinGB}
\bR_{MN}=\a\,\bH_{MN} -\La_D\,\bg_{MN} -
\frac{1}{D-2}\,\bg_{MN}\,\left(\a\,\bH - D\,\La_D \right). \ee
Using (\ref{0EinsteinGB}) together with (\ref{PRicci}) the first
order field equations can be written as \be \lb{ModifiedLichne}
\Box\,h_{MN} - 2 \bR_{KMNL}\,h^{KL} =  \a\,\mathcal{B}_{MN}, \ee
where \bea
\mathcal{B}_{MN} = -2\d H_{MN} + \frac{2}{D-2}\,\bg_{MN}\,\bg^{KL}\d H_{KL} - \frac{2}{D-2}\,\bg_{MN}\,\bH_{KL}\,h^{KL} \nn \\
+\bH^K_M\,h_{KN} + \bH^K_N\,h_{MK}. \eea

%{\bf According to my notes $\delta H = \bar g^{KL}\delta H_{KL}$.
%\underline{Vassili}, can you check this equation? I remember there
%was an additional simplification, I send you the notes I have.}
%{\bf \underline{Bertha}, I agree that $\delta H = \bar g^{KL}\delta H_{KL}$.
%but finally there is not some simplification according to my notes. There was some for when we took the wrong definition
%$\delta H = \delta (H^{MN} g_{MN})$. If you finally agree with me just erase those comments.}

When we calculated the variation of the Gauss-Bonnet term,
the harmonic gauge was used. However, there is still some gauge freedom
which can be further gauged away.

The second term can be written as

\bea \lb{simp2}
\bg^{KL}\d H_{KL}=\left(2-D\right) h^{MN} \bR_{MN}
\bR +\left(8-2D\right) \bR^{MN} \Box\,h_{MN}
+\left(4D-10\right) \bR^{MS} \bR^{P\,\,\,\,N}_{\,\,M\,\,\,\,S} h_{PN}\nn \\
-\left(D-2\right) \bR^{ABMS} \bR^{P}_{\,\,BMS} h_{AP}+4
\bR^{LP}\bR^{T}_{\,\,P}h_{LT}+\left(2D-8\right) \bR^{ABMS}
\nabla_{M} \nabla_{B}h_{AS}\nonumber \\
\eea
while the third term becomes
\bea \lb{simp3}
\bH_{KL}\,h^{KL} =-
\bR\,\bR_{KL}\,h^{KL}+4\bR_{KP}\,\bR_{L}^{\,\,\,P}\,h^{KL}+4\bR^{M}_{\,\,\,KPL}\,\bR_{M}^{\,\,\,P}\,h^{KL}-2\bR_{KMNP}
R_{L}^{\,\,\,MNP}\,h^{KL}\nonumber \\
\eea

Equation (\ref{ModifiedLichne}) is the \emph{modified
Lichnerowicz equation}, the difference with (\ref{Lichne}) is
the appearance of the source term $\a\,\mathcal{B}_{MN}$ due to
the Gauss-Bonnet term in the action.

Another way of arriving to the same result is to vary directly
the equation (\ref{gbeq}). However,  the modified Lichnerowicz
equation (\ref{ModifiedLichne}) is useful because it gives
directly the perturbation equations once the form of the
perturbation is known.

\section{Perturbation analysis of spherically symmetric Gauss-Bonnet black hole solutions}

As an application of the formalism developed in the previous
section we will examine the perturbations of spherically symmetric
solutions of Einstein-Gauss-Bonnet equations
\be\lb{gbeq2}
G_{MN}-\a H_{MN}=-\La_D\,g_{MN},
\ee
where $G_{MN}$ is the Einstein tensor and $H_{MN}$ is the Gauss-Bonnet tensor
\bea \label{GB}
H_{M}^{\,\,N}&=&\frac{1}{2}g_{M}^{\,\,N}\left(R^2-4R_{KL}\,R^{KL}+R_{KLPQ}R^{KLPQ}\right)-
2R\,R_{M}^{\,\,N} \\ \nn &\,&+ 4 R_{MK}\,R^{NK}+
4R_{KMP}^{\,\,\,\,\,\,\,\,\,\,\,\,\,\,\,\,\,N}\,R^{KP}-
2R_{MKLP}\,R^{NKLP}.
\eea
Namely, we will investigate the following
equation \be\lb{egbperteq} \d G_{A}^{\,\,B}-\a \d
H_{A}^{\,\,B}=-\La_{D}\d g_{A}^{\,\,B}. \ee

In the transverse traceless gauge the variation of the Einstein
tensor gives \be \d G_{AB}=-\frac{1}{2}\left(\square
h_{AB}+2R^{L\,\,\,K}_{\,\,\,A\,\,\,B}
h_{LK}-R^{L}_{\,\,A}h_{BL}-R^{L}_{\,\,B}h_{AL}\right)
-\frac{1}{2}h_{AB} R+\frac{1}{2}g_{AB}h^{KL}R_{KL} \ee and
\be\lb{EinstTenPert} \d G_{A}^{\,\,B}=g^{BK}\d
G_{AK}-h^{BK}G_{AK}. \ee
Furthermore,
\be\lb{GBTenPert}
\d H_{A}^{\,\,B}=g^{BK}\d H_{AK}-h^{BK}H_{AK}, \ee where the
variations $\d H_{AK}$ are given by the  equations
(\ref{PH1})-(\ref{PH5}) which already incorporate the transverse
and traceless gauge. Additionally, it is easy to see that \be \d
g_{A}^{\,\,B}=g^{BK} h_{AK}-h^{BK}g_{AK}=0. \ee

We will consider spherically symmetric black hole solutions of the
Einstein-Gauss-Bonnet theory in   $D=6$ dimensions. The procedure
is the same in the case of five dimensions. The resulting
equations are \be \d G^{\prime\,\,B}_{A}+\a^{\prime}\, \d
H^{\prime\,\,B}_{A}=0,  \ee where $\d H^{\prime\,\,B}_{A}$ are
given by (\ref{PH1})-(\ref{PH5}) and they coincide with the
equations derived in \cite{Dotti:2004sh} for tensor perturbations
with the identifications $-\frac{1}{2}H_{AB}=H^{\prime}_{AB}$ and
$\a=2\a^{\prime}$.

Following \cite{Dotti:2004sh} these equations can be solved
transforming them to a Schr\"odinger-like equation. Consider
 the following spherically symmetric metric
\begin{equation} \label{bh}
ds^2 = -f(r) dt^2 + \frac{1}{f(r)} dr^2 + r^2
\left[d\theta^{2}+\sin^{2}\theta\left(d\varphi^{2}+\sin^{2}\varphi\left(d\chi^{2}+\sin^{2}\chi
d\psi^{2}\right)\right)\right]
\end{equation}
which satisfies equations (\ref{gbeq2}). Perturbations of the
above metric read \cite{Dotti:2004sh}
\begin{equation}
g_{\a\b} \to g_{\a\b} + h_{\a\b},
\end{equation}
where
\be \lb{GBtensorperts}
h_{\a\b}=r^{2}\phi\left(r,t\right)\left( \begin{array}{cccccc}
0 & 0 & 0 & 0 & 0 & 0 \\
0 & 0 & 0 & 0 & 0 & 0 \\
0 & 0 & h_{\theta\theta} & h_{\theta\varphi} & h_{\theta\chi} & h_{\theta\psi} \\
0 & 0 & h_{\theta\varphi} & h_{\varphi\varphi} & h_{\varphi\chi} & h_{\varphi\psi} \\
0 & 0 & h_{\chi\theta} & h_{\chi\varphi} & h_{\chi\chi} & h_{\chi\psi} \\
0 & 0 & h_{\psi\theta} & h_{\psi\varphi} & h_{\psi\chi} &
h_{\psi\psi}  \end{array} \right). \ee For every mode in the
perturbation we have $h_{ij}=h_{ij}\left( \theta,\varphi,\chi,\psi
\right)$ with $i,j=\theta,\varphi,\chi,\psi$. Furthermore, the
transverse and traceless choice for the gauge alongside with the
symmetries of the metric implies  that the restriction of
$h_{\a\b}$ to the sphere is transverse, traceless, and so it can
be expanded using a basis of eigentensors of the Laplacian on the
sphere \cite{higu}. So with $i,j$ running on the sphere we have
\be h_{ij}=r^{2}\phi\left(r,t\right) \bar h_{ij}(x), \ee where \be
\bar \nabla^k \bar \nabla_k  \bar h_{ij} = \g \bar h_{ij}
,\hspace{1cm} \bar \nabla^i  \bar h_{ij} = 0 , \hspace{1cm} \bar g
^{ij}  \bar h_{ij} = 0.
\end{equation}
The bar refers to metric and tensors on the $S^{4}$.

 Substituting (\ref{GBtensorperts}) into (\ref{EinstTenPert})
  and using the expansion of $h_{ij}$ according to the basis of
  eigentensors of the Laplacian on the sphere we find equation $(11)$ of
  \cite{Dotti:2004sh}. But somehow this is expected.
  In addition, we find the same equations as
  in \cite{Dotti:2004sh} for the Gauss-Bonnet combination.
  Substituting again (\ref{GBtensorperts}) to (\ref{GBTenPert})
  and using both the expansion of $h_{ij}$ according to the basis
  of eigentensors of the Laplacian on the sphere, alongside with the
  transverse and traceless gauge, we end up after a lot of algebra to
  the equation $(12)$ of \cite{Dotti:2004sh}.

 The crucial point in the above analysis comes from the symmetries of spacetime.
  Here spherical symmetry
  allows for the decomposition of $h_{ij}$ according to the basis of eigentensors
  of the Laplacian on the sphere, which at the end is the key factor for obtaining
   the final master equation of the perturbation analysis. Such a simplification
   is more difficult and  some times not possible  in the cases where the spacetime allows for symmetries other
   than spherical as we will discuss in the following sections.

%%%%%%%%%%%%%%%%%%%%%%%%%%%%%%%%%%%%%%%%%%%%%%%%%%%%%%%%%%%%%%%%%%%%%%%%%%

\section{Perturbations of codimension-2 black strings}

We consider the following gravitational action in five dimensions
with a Gauss-Bonnet term in the bulk
\begin{eqnarray}\label{AcGBIG}
S_{\rm grav}&=&\frac{M^{3}_{5}}{2}\left\{ \int d^5 x\sqrt{-
g^{(5)}}\left[ R^{(5)}
%-2\Lambda_5
+\alpha\left( R^{(5)2}-4 R^{(5)}_{MN}R^{(5)MN}+
R^{(5)}_{MNKL}R^{(5)MNKL}\right)\right] \right.~.
%\nonumber\\
%&+& \left. r^{2}_{c} \int d^3x\sqrt{- g^{(3)}}\,R^{(3)}
%\right\}+\int d^5 x \mathcal{L}_{bulk}+\int d^3 x
%\mathcal{L}_{brane}\,,
\label{5daction}\nonumber\\
%\left[-\sigma + \frac{r}{2\kappa^2_5} R^{(4)}\right],
\end{eqnarray}
%where $\alpha\, (\geq0)$ is the Gauss-Bonnet coupling constant.
We are looking for solutions of the form \be
ds_5^2=g_{\m\n}(x,\rho)dx^\m
dx^\n+a^{2}(x,\rho)d\rho^2+L^2(x,\rho)d\th^2~.\label{5dmetric} \ee
The topology of the two-dimensional space $(\rho,\th) $ can be
represented by a cone of deficit angle $\b$. Regularization of
this space dictates the introduction of a brane  located at the
tip of the cone. As we will discuss, the presence of the brane has
important consequences in the perturbative analysis of these
spaces. Solutions of the form (\ref{5dmetric}) have been obtained
in \cite{CuadrosMelgar:2007jx}. In this section we will derive
the general formalism of metric perturbations of these solutions
 and we will discuss the scalar, vector, and tensor
perturbations.

 \subsection{General formalism of metric perturbations}

Consider the metric
 \be \lb{CPTZ}
ds^{2}=f^2(\rho)\left[ -n^2(r) dt^2 + \frac{dr^2}{n^2(r)} + r^2 d\phi^2\right] + d\rho^2 + b^2(\rho)d\theta^2 \,,
\ee
%\bea \lb{CPTZ}
%ds^{2}=\cosh^{2}\left(\frac{\r}{2 \sqrt{\a}}\right)\left[-\left(-M+\frac{r^{2}}{l^{2}}\right)\,dt^{2}+\left(-M+\frac{r^{2}}{l^{2}}\right)^{-1}\,dr^{2}+r^{2}d\phi^{2}\right] \nn \\
%+d\r^{2}+\left(2\,\b\,\sqrt{\a}\,\sinh\left(\frac{\r}{2
%\sqrt{\a}}\right)\right)^{2}d\theta^{2},
%\eea
which is of the form (\ref{5dmetric}). In what follows we will consider $f(\rho)=\cosh\left(\r/2 \sqrt{\a}\right)$,
$b(\r)=2\,\b\,\sqrt{\a}\,\sinh\left(\r/2\sqrt{\a}\right)$, and
$n^2(r)=-M+r^2/l^2$, which is a solution of the action
(\ref{5daction}) with the inclusion of a brane boundary term
 \cite{CuadrosMelgar:2007jx}.

In order to investigate the stability of the solution we need to
adopt an Ansatz for our perturbations. It is possible to see that
the above metric has $3$ Killing vectors. We can adopt a Fourier
expansion of the perturbation with respect to the coordinates
$t,\phi,\theta$ and keep all the components of the perturbation as
functions of $r,\r$ . We are going to consider perturbations which
retain the angular symmetry on the brane as well as on the
transverse space. Keeping the axial symmetry both on the brane and
in the transverse space means that everything must be Lie derived
by $\partial_{\phi}$ and $\partial_{\theta}$  to zero
\cite{Kaloper:2007ap}. In this way, we will work on the $s-wave$
approximation of our system for the two angular coordinates.

As a result of the symmetries of the spacetime we can split the
perturbation into a purely two-dimensional transverse piece, a
mixed transverse and three-dimensional piece and a purely
three-dimensional piece. This can be represented by \be
\lb{PertsDecom} \left( \begin{array}{cc}
h_{\mu \nu} & h_{\mu i} \\
h_{j \nu} & h_{ij}  \end{array} \right) \ee  where
$(\m,\n=t,r,\phi)$ and $(i,j=\r, \theta ) $. In the Kaluza-Klein
spirit, these perturbations can be interpreted as scalar, vector,
and tensor, respectively, with respect to the three-dimensional
spacetime. We will consider the following  Ansatz for our
perturbation

\be \lb{PertsAns1} h_{AB}=e^{\Omega t}\left( \begin{array}{cc}
h_{\mu \nu}\left(r,\r\right) & h_{\mu i}\left(r,\r\right) \\
h_{j \nu}\left(r,\r\right) & h_{ij}\left(r,\r\right) \end{array}
\right) \ee
% \be
%\lb{PertsAns1} h_{AB}=e^{\Omega t}\left(
%\begin{array}{ccccc}
%h_{tt}\left(r,\r\right) & h_{tr}\left(r,\r\right) & h_{t\phi}\left(r,\r\right) & h_{t\r}\left(r,\r\right) & h_{t\theta}\left(r,\r\right) \\
%h_{tr}\left(r,\r\right) & h_{rr}\left(r,\r\right) & h_{r\phi}\left(r,\r\right) & h_{r\r}\left(r,\r\right) & h_{r\theta}\left(r,\r\right) \\
%h_{t\phi}\left(r,\r\right) & h_{r\phi}\left(r,\r\right) & h_{\phi\phi}\left(r,\r\right) & h_{\r\phi}\left(r,\r\right) & h_{\phi\theta}\left(r,\r\right) \\
%h_{t\r}\left(r,\r\right) & h_{r\r}\left(r,\r\right) & h_{\r\phi}\left(r,\r\right) & h_{\r\r}\left(r,\r\right) & h_{\r\theta}\left(r,\r\right) \\
%h_{t\theta}\left(r,\r\right) & h_{r\theta}\left(r,\r\right) &
%h_{\phi\theta}\left(r,\r\right) & h_{\r\theta}\left(r,\r\right) &
%h_{\theta\theta}\left(r,\r\right) \end{array} \right) \ee

The above ansatz contains  the maximum information concerning the
perturbation of our system, while at the same time is consistent
with the aforementioned arguments.

Substituting (\ref{PertsAns1}) into the Lichnerowicz equation
(\ref{ModifiedLichne}), we can see in a straight forward manner
that, the $h_{\phi\theta}(r,\rho)$ mode decouples from the rest of
the modes straight away, while the $h_{t\phi}\left(r,\r\right)$,
$h_{r\phi}\left(r,\r\right)$ and $h_{\r\phi}\left(r,\r\right)$
modes are coupled together and the same  happens also for
$h_{t\theta}\left(r,\r\right)$, $h_{r\theta}\left(r,\r\right)$ and
$h_{\r\theta}\left(r,\r\right)$ modes. We could set them to zero
in order to examine the rest of the system, but for the moment we
will keep them, since in any case these modes, for the s-wave
approximation that we are considering, do not interact with the
other modes. As a first step we will examine the behaviour of the
scalar modes $h_{\theta\theta}\left(r,\r\right)$,
$h_{\r\r}\left(r,\r\right)$ and $h_{\r\theta}\left(r,\r\right)$.

\subsection{Scalar perturbations-solving the Lichnerowicz  equation}

We consider first the scalar perturbations. Our aim is to solve
the modified Lichnerowicz  equation (\ref{ModifiedLichne}) in the
background metric (\ref{CPTZ}) using the perturbation Ansatz
(\ref{PertsAns1}). Substituting our Ansatz for the perturbation we can see that all the perturbation equations are governed by a prefactor,
 which has the form $\left(l^{2}-4\a\right)$. We see that when
 we are close to $l^{2}\to4\a$, we face a strong coupling problem.
 We will refer to this limit later on.

Using (\ref{simp2}) and (\ref{simp3}) we can
calculate the two equations for the scalar modes. These equations
couple the scalar modes of the perturbation with the tensor and
the vector modes. However, they can be decoupled. Here we will
summarize the results while in the appendix we present the
technical details. The two scalar modes are given by

\bea
h_{\r\r}&=&\frac{1}{\b^{2}}\partial_{\r}\left[\frac{h_{\theta\theta}}{\sqrt{a}
\sinh\left(\frac{\r}{\sqrt{\a}}\right)}\right]\lb{SecondScalarEq},\\
h_{\theta\theta}&=&\left[2\,\b\,\sqrt{\a}\,\sinh\left(\frac{\r}{2
\sqrt{\a}}\right)\right]^{2}
u\left(r,\r\right).\lb{hthetatheta}\eea

Separating the variables choosing
$u\left(r,\r\right)=f\left(r\right)\,y\left(\r\right)$, the
functions $f\left(r\right)$ and $y\left(\r\right)$ are given by
the differential equations
\be \lb{KKmodesEq1}
\frac{4\,\a^{3/2\,}coth\left(\frac{\r}{2\,\sqrt{\a}}\right)}{l^{2}}\,\left[3\,\cosh\left(\frac{\r}{\sqrt{\a}}\right)\,\frac{\partial\,y\left(\r\right)}{\partial\,\r}+\sqrt{\a}\,\sinh\left(\frac{\r}{\,\sqrt{\a}}\right)\,\frac{\partial^{2}\,y\left(\r\right)}{\partial\,\r^{2}}
\right]+m\,y\left(\r\right)=0,
\ee
with $m$ being the separation constant
\be \lb{FinalScalar1Eq1}
\left(m-\frac{8\,\a^{2}\,\Omega^{2}}{l^{2}\,M-r^{2}}\right)\,f\left(r\right)+\frac{8\,
\a^{2}}{l^{4}\,r}\left[\left(l^{2}\,M-3r^{2}\right)\frac{\partial\,f\left(r\right)}
{\partial\,r}+r\left(l^{2}\,M-r^{2}\right)\,\frac{\partial^{2}\,f\left(r\right)}{\partial\,r^{2}}\right]=0.
\ee

The equation (\ref{FinalScalar1Eq1}) can be written  in the
following form
\be \lb{FinalScalar2Eq}
\left(-l^{2}\,M+r^{2}\right)\,\frac{\partial^{2}\,f\left(r\right)}{\partial\,r^{2}}+\frac{\left(-l^{2}\,M+3r^{2}\right)}{r}\frac{\partial\,f\left(r\right)}{\partial\,r}+\frac{l^{4}}{8
\a^{2}}\left(\frac{8\,\a^{2}\,\Omega^{2}}{l^{2}\,M-r^{2}}-m\right)\,f\left(r\right)=0.
\ee
Setting
$f\left(r\right)=\frac{\psi\left(r\right)}{\sqrt{r}}$, and
introducing the tortoise coordinate defined by
$$dr_{*}=\frac{dr}{-M+\frac{r^{2}}{l^{2}}}$$
Eq.(\ref{FinalScalar2Eq}) can be written as a Schr\"odinger
equation \be \lb{SchrondingerEq}
\frac{d^{2}\psi\left(r\right)}{dr_{*}^{2}}+\left[\,V\left(r\right)-\Omega^2\right]\,\psi\left(r\right)=0,
\ee where $V\left(r\right)$ reads \be \lb{Potential}
V\left(r\right)=\frac{M}{2l^{2}}+\frac{M^{2}}{4r^{2}}-\frac{3r^{2}}{4l^{4}}+\frac{l^{2}mM}{8\a^{2}}-\frac{mr^{2}}{8\a^{2}}\,.
\ee Note that the above potential depends only on the separation
constant and the Gauss-Bonnet coupling constant $ \alpha$. This is
the only information that it has from the bulk. It does not depend
on the deficit angle $\beta$. This can be understood because
$b(\rho)$ expresses the response of the bulk geometry to the
presence of the conical singularity, information that is entirely
encoded in the decoupled equation (\ref{KKmodesEq1}), which
describes the behaviour of the perturbation in the
extra-dimensions. We will show in the following subsection that
the deficit angle explicitly appears in this extra-dimensional
equation only if we consider an angular dependence in the total
perturbation as can be seen in Eq.(\ref{rho}) together with its
solution (\ref{bsolution}).

 The above potential is similar to the one found in the
calculation of quasinormal modes of BTZ black hole
\cite{Cardoso:2001hn}. Furthermore, it is exactly the same  as the
potential we will find in the next subsection by solving the
Klein-Gordon equation for the background metric, provided that we
make the identification $\n^{2} \to \frac{l^{2}m}{8\a^{2}}$. The
Schr\"odinger equation (\ref{SchrondingerEq}) will be solved in
 the next subsection. The solution  implies that \be\label{omega-scM}
\Omega=-\frac{\sqrt{M}}{l}\left(1+2N+\sqrt{1+\frac{ml^{4}}{8
\a^{2}}}\right)\,, \ee where $N$ is a positive integer. As this
quantity is always negative, it yields a decaying solution for the
perturbation assuring the stability of the model under scalar
metric perturbations. Although in general the mass may also take
negative values,
%We can see that $\Omega$ is always negative, even if $m$ takes negative values until the limit $-\frac{8\a^{2}}{l^{4}}$. This limit is directly connected with the Breitenlohner-Freedman limit \cite{Breitenlohner:1982jf},\cite{Breitenlohner:1982bm}.}
the spectrum of $m$ can be analysed by looking at the bulk equation (\ref{KKmodesEq1}), whose general solution is
\be \label{hypergeom1}
y(\r)={\cal C}_1\,q^{A}\, _2F_1(A,B;C;q) + {\cal C}_2\,q^{A'}\, _2F_1(A',B';C';q)\,,
\ee
with
\bea
q&=&\cosh^2(\r/2\sqrt\a)\,,\\
A&=&- \frac{1}{2}-\frac{1}{4}\sqrt{4+2l^2m/\a}\,,\\
B&=&\frac{5}{2} - \frac{1}{4}\sqrt{4+2l^2m/\a}\,,\\
C&=&1 - \frac{1}{2}\sqrt{4+2l^2m/\a}\,,\\
A'&=&-C/2\,,\\
B'&=&B-A-C/2\,,\\
C'&=&-2A\,. \eea For this solution let us choose ${\cal C}_1$ and
${\cal C}_2$ such that we can obtain a decaying behaviour which
ascertains a smooth transition towards infinity. As the arguments of
the hypergeometric functions need to be real, the first
restriction on $m$ comes from the square root: $m\ge -2\a/l^2$.
The equality in this equation when substituted in
(\ref{omega-scM}) corresponds to the strong coupling or
Chern-Simons limit. Interesting enough, at this limit the extra
term disappears and we recover the results of a three-dimensional
quasinormal modes of the BTZ black hole \cite{Cardoso:2001hn}.
However, the hypergeometric functions are not well-defined for all
the values of $m$ above this limit. This strictly depends upon the
parameters of each function. A careful analysis shows that ${\cal
C}_2$ needs to be zero since it cannot produce decaying solutions.
Moreover, recalling the properties of these functions, we were
able to find a special case by setting $B=0$, which corresponds to
$m=48\a/l^2$, where the perturbation manifests a decaying
behaviour.

We have one more scalar mode namely the $h_{\rho\theta}$ mode.  As
we have already mentioned, this mode is coupled with the
$h_{t\theta}$ and $h_{r\theta}$ modes. Still using the transverse
gauge, it can be easily seen, that this scalar mode also decouples
from the other two modes and quite surprisingly it obeys the same
Schr\"odinger equation as the $h_{\theta\theta}$ mode.

\subsection{Scalar perturbations-solving the Klein-Gordon equation}

In this subsection we will carry out the analysis of the scalar
perturbations solving the   Klein-Gordon equation. We will show
that we get the same results as the ones we  got by solving the
Lichnerowicz equation, but the study of the Klein-Gordon equation
gives us a better understanding of the stability of the
codimension-2 black string.

 Consider the massive Klein-Gordon equation
\begin{equation}
\Box \Phi = \frac{1}{\sqrt{-g}}\partial_M (\sqrt{-g} g^{MN}
\partial_N) \Phi = m^2 \Phi \,.
\end{equation}
By using the metric (\ref{CPTZ}) and decomposing the scalar field as
\begin{equation}
\Phi (t,r,\phi, \rho, \theta) = Z(t,r,\rho) \Xi (\phi) \Theta
(\theta)\,,
\end{equation}
we arrive to the following equation,
\begin{eqnarray}
-\frac{1}{f^2 n^2} \frac{\partial_t ^2 Z}{Z} + \left( \frac{n^2}{f^2 r} + \frac{2n\dot n}{f^2}\right) \frac{\partial_r Z}{Z} + \frac{n^2}{f^2} \frac{\partial_r ^2 Z}{Z} + \left(\frac{3f'}{f} + \frac{b'}{b}\right) \frac{\partial_\rho Z}{Z} + \frac{\partial_\rho ^2 Z}{Z} &&\nonumber \\
+\frac{l^2}{f^2 r^2} \frac{\partial_\phi ^2 \Xi}{\Xi} +
\frac{1}{b^2}\frac{\partial_\theta ^2 \Theta}{\Theta} - m^2 = 0
\,.
\end{eqnarray}
We can apply variable separation method to the $\theta$- and
$\phi$-dependent parts of this equation. Thus, the solutions to
$\Theta(\theta)$ and $\Xi(\phi)$ are given by
\begin{eqnarray}
\Theta(\theta) &=& A e^{i\kappa \theta} + B e^{-i\kappa \theta} \\
\Xi (\phi) &=& C e^{i\epsilon\phi} + D e^{-i\epsilon\phi}\,,
\end{eqnarray}
where $A$, $B$, $C$, and $D$ are integration constants while
$\kappa$ and $\epsilon$ are constants generated through the
variable separation. The remaining equation can be decoupled in
two equations with the Ansatz $Z(t,r,\rho)=\Psi(t,r) P(\rho)$ as
follows,
\begin{eqnarray}
\partial_t ^2 \Psi - n^2 \left(\frac{n^2}{r}+ 2n\dot n\right) \partial_r \Psi - n^4 \partial_r ^2 \Psi
+ \left( \frac{\epsilon^2 l^2 n^2}{r^2} + \nu ^2 n^2\right) \Psi &=& 0 \label{tr}\\
\partial_\rho ^2 P + \left( 3\frac{f'}{f} + \frac{b'}{b} \right) \partial_\rho P + \left( -\frac{\kappa^2}{b^2} - m^2
+ \frac{\nu^2}{f^2} \right) P &=& 0 \,, \label{rho}
\end{eqnarray}
being $\nu^2$ the constant of variable separation. Equation
(\ref{tr}) describes the evolution of the scalar perturbation on
the brane, while equation (\ref{rho}) says how this perturbation
behaves at different distances away from the brane.

In order to solve Eq.(\ref{tr}) we should rewrite the equation in
terms of the  tortoise coordinate,
\begin{equation}\label{tortuga}
r_* = \int \frac{dr}{n^2} \,.
\end{equation}
Thus, Eq.(\ref{tr}) adopts the Schr\"odinger form,
\begin{equation}\label{schro}
-\frac{\partial^2 X (t,r_*)}{\partial t^2} + \frac{\partial^2 X
(t,r_*)}{\partial r_* ^2} = V[r(r_*)] X (t,r_*) \,.
\end{equation}
Here $X (t,r_*)$ is a new function defined as
\begin{equation}
X (t,r_*)= \sqrt r\, \Psi(t,r)\,,
\end{equation}
and the so-called potential term can be written as
\begin{equation}\label{pot}
V(r) = \frac{n^2}{2r} \left( 2n \dot n - \frac{n^2}{2r} \right) +
\frac{\epsilon ^2 l^2 n^2}{r^2} + \nu^2 n^2 \,.
\end{equation}
When we compare this potential with the pure 3-dimensional case
studied in~\cite{Cardoso:2001hn}, we see that Eq.(\ref{pot}) has a
$\nu^2 n^2$ additional correction. This new term connects brane to bulk
perturbations through the constant $\n$ and does not vanish, in general.

We choose the time dependence as $e^{-i\omega t}$, so that
$X(t,r)=e^{-i\omega t} R(r)$, thus, Eq.(\ref{schro}) becomes
\begin{equation}\label{schro2}
\frac{d^2 R}{dr_* ^2} + [\omega^2 -V(r)] R =0\,.
\end{equation}

When $n(r)$ is of the BTZ form,  $n^2(r)=-M+r^2/l^2$, the potential (\ref{pot}) becomes
\begin{equation}
V(r) = \left(\frac{3}{4l^4} + \frac{\nu^2}{l^2} \right) r^2 -
\left( \frac{M}{2l^2} + M\nu^2 -\epsilon^2 \right) +
\left(-\frac{M^2}{4} - l^2 M\epsilon^2 \right) \frac{1}{r^2} \,.
\end{equation}
The tortoise coordinate can analytically be found from
(\ref{tortuga}) as
\begin{equation}
r = -l\sqrt M \,\coth\left( \frac{r_* \sqrt M}{l}\right)\,.
\end{equation}
Notice that $r_H=l\sqrt M \leq r < \infty$ corresponds to $-\infty
< r_* \leq 0$.

Let us make the following variable transformation,
\begin{equation}
x = \frac{1}{\cosh ^2 (\frac{r_* \sqrt M}{l})}\,,
\end{equation}
with $0\leq x\leq1$, so that the potential turns into
\begin{equation}
V(x) = -\frac{x}{4} \left[\frac{4M(1+\nu^2 l^2)-Mx
-4l^2 \epsilon^2(x-1)}{l^2 (x-1)}\right]
\end{equation}
and Eq.(\ref{schro2}) becomes
\begin{equation}\label{schro3}
4x(1-x) \frac{d^2 R}{dx^2} + (4-6x) \frac{dR}{dx} + \left[
\frac{\omega^2 l^2}{Mx}+\frac{1+\nu^2 l^2}{x-1} - \frac{x}{4(x-1)}
-\frac{l^2 \epsilon^2}{M} \right] R =0\,.
\end{equation}

In order to write this equation in a more familiar way we make a
new substitution as follows,
\begin{equation}
R = \frac{(x-1)^{3/4}}{x^{i\omega l/2\sqrt M}}\, y(x)\,.
\end{equation}
Then, Eq.(\ref{schro3}) turns to be
\begin{equation}
-x(x-1)y'' + \left[1-3x+\frac{i\omega l}{\sqrt M}(x-1) \right] y' +
\left[ \frac{\omega^2 l^2}{4M} + \frac{i\omega l}{\sqrt M} -
\frac{l^2 \epsilon^2}{4M} -1 + \frac{\nu^2 l^2}{4(x-1)} \right] y =0\,.
\end{equation}

The general solution of this equation can be expressed in terms of
hypergeometric functions of the second kind $_2 F_1 (a,b;c;x)$. As
we look for stable behaviour, we keep the decaying branch of the
general solution and establish the restrictions for this
stability,
\begin{equation}\label{sol1}
y(x) = (1-x)^{-(1+\sqrt{1+\nu^2 l^2})/2} \phantom{A}_2 F_1
(a,b;c;x)\,,
\end{equation}
where
\begin{eqnarray}\label{abc}
a &=& \frac{1}{2} \left( 1 - \sqrt{1+\nu^2 l^2} - \frac{i\omega l}{\sqrt M} + \frac{i l\epsilon}{\sqrt M} \right) \nonumber \\
b &=& \frac{1}{2} \left( 1 - \sqrt{1+\nu^2 l^2} - \frac{i\omega l}{\sqrt M} - \frac{i l\epsilon}{\sqrt M} \right) \nonumber \\
c &=& 1-\frac{i\omega l}{\sqrt M}\,.
\end{eqnarray}

Since the (2+1) spacetime is asymptotically AdS, the correct
boundary condition to be taken in Eq.(\ref{sol1}) is the flux
condition, {\it i.e.},
\begin{equation}\label{flujo}
{\cal F} \sim (y^* \partial_\mu y - y \partial_\mu
y^*)\Big|_{x=1}=0 \,.
\end{equation}
%This implies
%\begin{equation}\label{flux}
%_2F_1 ^* (a,b;c;x=1)\; _2F_1 \,^\prime(a,b;c;x=1) -\, _2F_1 (a,b;c;x=1)\; _2F_1 ^* \,^\prime(a,b;c;x=1) =0\,.
%\end{equation}
In order to evaluate this condition at $x=1$ we can use the following property of hypergeometric functions.
Given a $_2 F_1 (a',b';c';x)$, if $c'$ is not a negative integer,
the series converges when $x=1$ if $\Re (c'-a'-b')>0$, and we can
write the hypergeometric function as
\begin{equation}\label{glux}
_2 F_1 (a',b';c';1) = \frac{\Gamma(c')
\Gamma(c'-a'-b')}{\Gamma(c'-a') \Gamma(c'-b')}\,.
\end{equation}
%Applying this property to the first term in (\ref{flux}) and choosing the derivative to vanish we have, \footnote{This choosing proves to be the precise condition to guarantee the stability of the solution as we will see.}
The derivative of the hypergeometric functions we are dealing with has the following form,
\begin{equation}\label{fluxder}
_2F_1 \,^\prime(a,b;c;x=1) = \frac{ab}{c} \,_2F_1(a+1,b+1;c+1;x=1) \,.
\end{equation}
Using (\ref{abc}) we verify that $\Re (c-a-b-1)=\sqrt{1+\nu^2 l^2} -1>0$
since $\nu^2>0$. Thus, we can write (\ref{fluxder}) in terms of $\Gamma$ functions as in (\ref{glux}) and put back in the flux condition (\ref{flujo}). The new equation is fulfilled when $c-a=-N$ or $c-b=-N$, where $N$ is a positive integer. In this way we obtain the quasinormal frequencies,
\begin{equation}\label{qnf}
\omega = \pm \epsilon -i\frac{\sqrt M}{l} (1+2N + \sqrt{1+\nu^2 l^2}) \,.
\end{equation}
%This $\omega$ also implies $_2F_1(a,b;c;x=1)=0$, so that the second term in (\ref{flux}) vanishes automatically fulfilling the entire equation.
We observe that the imaginary part of these frequencies is
negative, thus, displaying the stability of the model under scalar
perturbations. Notice that in (\ref{qnf}) a term proportional to
$\nu^2$ appears which encodes the information from the bulk and
when $\nu^2=0$ the  pure three-dimensional BTZ
results~\cite{Cardoso:2001hn} are recovered.

To complete our analysis we must solve Eq.(\ref{rho}). We have two
subcases.
\begin{enumerate}
\item When $f(\rho)=\cosh (\rho/2\sqrt \alpha)$ and
$b(\rho)=2\beta\sqrt\alpha \sinh(\rho/2\sqrt \alpha)$, the most general
solution of Eq.(\ref{rho}) is given by
\begin{equation}
P(z) = \frac{(z-1)^{\kappa/2\beta}}{\sqrt{2z}} \left[ C_1
z^{-\sqrt{1+4\nu^2\alpha}/2} \phantom{A}_2 F_1 (\hat a, \hat b;
\hat c;z) + C_2 z^{\sqrt{1+4\nu^2\alpha}/2} \phantom{A}_2 F_1
(\hat a', \hat b'; \hat c';z)\right]\,, \label{bsolution}
\end{equation}
where
\begin{equation}
z = \cosh^2 \left(\frac{\rho}{2\sqrt\alpha} \right) \,,
\end{equation}
$C_1$ and $C_2$ are constants, and
\begin{eqnarray}
\hat a &=& \frac{1}{2} \left( 1 +\frac{\kappa}{\beta} - \sqrt{1+4\nu^2\alpha} + 2\sqrt{1+m^2\alpha}\right)\nonumber \\
\hat b &=& \frac{1}{2} \left( 1 +\frac{\kappa}{\beta} - \sqrt{1+4\nu^2\alpha} - 2\sqrt{1+m^2\alpha}\right)\nonumber \\
\hat c &=& 1-\sqrt{1+4\nu^2\alpha}\nonumber \\
\hat a' &=& \frac{1}{2} \left( 1 +\frac{\kappa}{\beta} + \sqrt{1+4\nu^2\alpha} - 2\sqrt{1+m^2\alpha}\right)\nonumber \\
\hat b' &=& \frac{1}{2} \left( 1 +\frac{\kappa}{\beta} + \sqrt{1+4\nu^2\alpha} + 2\sqrt{1+m^2\alpha}\right)\nonumber \\
\hat c' &=& 1+\sqrt{1+4\nu^2\alpha}\,.
\end{eqnarray}

A careful study of both hypergeometric functions shows that $C_2$
needs to be  zero if we want to have a decaying solution. Given a
suitable set of parameters the remaining hypergeometric function
has the pursued behaviour. A special case appears when $\nu ^2$
adopts the following form,
\begin{equation}
\nu^2 = m^2 +\frac{1}{\alpha} +\frac{\kappa}{2\alpha\beta}\left(1+\frac{\kappa}{2\beta} \right) + \frac{\sqrt{m^2\alpha+1}}{\alpha} \left(1+\frac{\kappa}{\beta} \right)\,,
\end{equation}
which produces $b=0$. This case corresponds to a damped
oscillation in $\rho$. Therefore, the model  is well behaved under
this kind of perturbation having the quasinormal frequencies given
in (\ref{qnf}).

\item When $f(\rho)=\pm 1$ and $b(\rho)=\gamma \sinh
(\rho/\gamma)$, with $\gamma=\sqrt{(l^2-4\alpha)/2}$, we find
the following solution for (\ref{rho}),
\begin{equation}
P(w) = \sqrt 2 (w-1)^{\kappa/2} \left[ C_3 \phantom{A}_2 F_1
(\tilde a, \tilde b; \tilde c; w) + C_4 \sqrt{2w} \phantom{A}_2
F_1 (\tilde a', \tilde b'; \tilde c'; w) \right]\,,
\end{equation}
where
\begin{equation}
w = \cosh^2 \left( \frac{\sqrt 2\, \rho}{\sqrt{l^2-4\alpha}}
\right) \,,
\end{equation}
$C_3$ and $C_4$ are constants, and
\begin{eqnarray}
\tilde a &=& \frac{1}{4} + \frac{\kappa}{2} - \frac{1}{4} \sqrt{1+(l^2-4\alpha)(2m^2-2\nu^2)} \nonumber \\
\tilde b &=& \frac{1}{4} + \frac{\kappa}{2} + \frac{1}{4} \sqrt{1+(l^2-4\alpha)(2m^2-2\nu^2)} \nonumber \\
\tilde c &=& \frac{1}{2} \nonumber \\
\tilde a' &=& \frac{3}{4} + \frac{\kappa}{2} + \frac{1}{4} \sqrt{1+(l^2-4\alpha)(2m^2-2\nu^2)} \nonumber \\
\tilde b' &=& \frac{3}{4} + \frac{\kappa}{2} - \frac{1}{4} \sqrt{1+(l^2-4\alpha)(2m^2-2\nu^2)} \nonumber \\
\tilde c' &=& \frac{3}{2} \,.
\end{eqnarray}

In this case both hypergeometric functions display a growing
behaviour.  This means that the scalar perturbation when going
into the bulk propagates without boundary, an analogous behaviour
to the one found in the five-dimensional black string solution of
\cite{Chamblin:1999by}.
\end{enumerate}

In the last two subsections  we studied the scalar perturbations
of the five-dimensional black string solution of codimension-2.
Using both the Lichnerowicz equation and the Klein-Gordon equation
we found stability under scalar perturbations.  For an alert
reader we  comment on the approximation we used.  We have chosen
the s-wave approximation for our ansatz, both for the angular
coordinate on the brane and on the transverse space. However, on
general grounds the  Fourier decomposition should also include,
the frequency modes of the two angular dimensions. Still it is not
clear to us whether the deficit angle is playing a significant
role in the stability analysis. The  procedure we have followed in
the quasi-normal mode analysis shows (see equations (\ref{tr}) and
(\ref{rho})) that each angular coordinate, affects the
corresponding spatial dimension. The $\phi$ dependence is
reflected in the equation for the brane, while the $\theta$
dependence and hence the deficit angle comes with the transverse
space equation.  Any effect of the deficit should be reflected on
the solution of the transverse piece. However, the only
information that equation (\ref{rho}) give us is to tell us how
the perturbation created on the brane is transmitted into the
bulk.

      Should we had chosen a frequency mode in the Fourier decomposition for the $\theta$ dimension,
      then we should examine whether the scalar part decouples from the rest of the system
      and even if the variables in the Klein-Gordon equation can be separated.
      This procedure is very  involved,  due to the presence of the Gauss-Bonnet's quadratic terms
      and it is beyond the scope of the present work.  This procedure may give different results,
      however, even in this case, the deficit angle due to its intrinsic geometrical nature,
      may still has an effect only on the transverse space.

%%%%%%%%%%%%%%%%%%%%%%%%%%%%%%%%%%%%%%%%%%%%%%%%%%%%%%%%%%%%%%%%%%%%%%%%%%%

\subsection{Vector and tensor perturbations}

We have seen that  as far as we are away from the Chern-Simons
limit we can always apply the transverse and traceless gauge and
examine the stability behaviour. We saw that for scalar
perturbations the system declines exponentially, meaning that
$\Omega$ is negative, even though the mass of the scalar mode
takes negative values until a certain limit. This feature ensures
that our system is stable under scalar metric perturbations. Of
course, the quadratic nature of the Gauss-Bonnet combination does
not give straightforward decoupled equations for the scalar part
and the decoupling procedure is not easy, as it happens in the
case of having just the Einstein tensor where the scalar part of
the perturbation decouples immediately.

It is known that a category of spherically symmetric vacuum
solutions with a Gauss-Bonnet term suffer from ghost-like
instabilities, and furthermore, there is a strong coupling problem
close to the Chern-Simons limit \cite{Charmousis:2008ce}. In our
case we  expect   the strong coupling problem to be present also
in the vector and tensor perturbations. Our analysis has shown
that under the particular Ansatz that we have chosen for the
perturbation there is a proportionality factor to every equation,
namely, $(l^{2}-4\a)$. As $l^{2}\to4\a$, all equations are
identically satisfied. This means that the first order
parametrization of the perturbation is not significant, and we
have to move to the second order.
% From a different point of view
%even though there is no matter in our investigation, meaning that
%we are studying the stability of the solution away from sources,
%eventually all the modes will interact with some kind of matter.
%The matter content will appear at the r.h.s. of the pertubation
%equations. At the aforementioned limit we see that all equations
%blow up. This is a direct signal of the strong coupling problem.

 We saw at the beginning of this section that the
scalar part of our perturbation, can be decoupled from the rest of
the system, with the help of the transverse  and traceless gauge.
Furthermore, a careful investigation of the perturbation equations
unveiled that the scalar perturbations do not have any
pathological behaviour. To discuss the other kind of perturbations
we will set these modes equal to zero and examine the rest of the
system.

The equation for the vector mode $h_{\phi\theta}$ is very simple
and it is decoupled in a straight forward manner from the rest of
the system \be \label{vectoreqs} coth\left(\frac{\r}{
\sqrt{\a}}\right)\frac{\partial
h_{\phi\theta}}{\partial\r}-\sqrt{\a} \frac{\partial^{2}
h_{\phi\theta}}{\partial\r^{2}}=0. \ee The solution of this
equation is trivial and it gives information only about the
transverse space.  As we can see there is no information about the
brane. Before making a comment on that let us see the equation for
$h_{t\theta}$ and $h_{r\theta}$ modes. Making use of the
transverse gauge, these modes decouple from each other and the
result is that both of them obey the same equation as
$h_{\phi\theta}$, namely equation (\ref{vectoreqs}). This is quite
unnatural from many aspects. First of all since the brane
dependence is arbitrary, this means that it could be satisfied by
a mode that has positive $\Omega$, signaling this way departure
from stability. Furthermore, the system is degenerate since these
three modes behave exactly the same, while at the same time we can
not make any solid prediction about its brane behaviour. This
could be a signal of a strong coupling problem, but it is not
clear whether this is exactly the case or not.

We still have the tensor part plus one more vector.  It can be
seen that choosing any of the remaining perturbation equations,
with the help of the transverse and traceless gauge,  we can set
them identically equal to zero. This is again an unnatural result.
Either it means that we have no propagating degrees of freedom,
which indeed could be the case because in three dimensions we do
not have a graviton, or again we are in a strong coupling regime,
which restricts us from making any conclusion from the linear
terms and we have to move forward to higher order ones
\footnote{In \cite{Banados:2005rz} the holographic description of
the Gauss-Bonnet theory  in five dimensions was discussed and it
was found that  there is a particular Weyl anomaly that prevents
the Gauss-Bonnet theory to go smoothly to the Chern-Simons limit
 signaling the breakdown of perturbation theory at linear order.
}. However, the case where we have no propagating degrees of
freedom seems more probable, since in the s-wave approximation
that we have chosen,
 the brane can "see" the transverse space through the
angular symmetry that is manifested both on the brane and on the
transverse space. The absence of a graviton mode on the brane, due
to its three dimensional nature is reflected also in transverse
space due to the aforementioned symmetry.

\section{Conclusions}

 We studied the metric perturbations of a five-dimensional black
 string  in codimension-2 braneworlds with a Gauss-Bonnet term in
 the bulk. After reviewing the general formalism of linear metric
 perturbations, we derived the modified Lichnerowicz equation in
 the presence of the Gauss-Bonnet term. As an application, we considered the  six-dimensional spherically symmetric Gauss-Bonnet black hole
 solutions and we applied the modified Lichnerowicz equation to
  these  solutions to derive the known results for the tensor perturbations.

We considered  the scalar perturbations of a five-dimensional
black string solution of codimension-2 braneworlds with a
Gauss-Bonnet term. We carried out the analysis using both the
Lichnerowicz equation and solving explicitly the Klein-Gordon
equation. Away from the Chern-Simons limit, we showed that the
results from the two methods coincide. The behaviour of the black
string under scalar perturbations can be described by two
equations. One of them describes the evolution of the scalar
perturbation on the brane while the other equation shows how this
perturbation behaves at different distances from the brane. The
evolution of the black string on the brane is controlled by the
quasinormal modes which are similar to the quasinormal modes of
the BTZ black hole with the addition of an extra term which has
the information from the bulk. The stability analysis shows that
the black string is stable under scalar perturbations.

We also studied the vector and tensor perturbations of the black
string solution. We found that for the vector modes we can only
get information on their behaviour in the bulk while for the
tensor modes we did not find any physical propagating modes. We
mainly attribute this behaviour to the specific symmetries of the
considered black string solution but also may be an indication of
 a strong coupling problem signaling the  need to
go to the next order in perturbation theory.

It is interesting to extent the analysis to the six-dimensional
black string solutions derived in \cite{CuadrosMelgar:2008kn}. In
six dimensions there is also a limit where the theory becomes
strongly coupled. Away from this limit, we expect the application
of the modified Lichnerowicz equation to this solution to simplify
significantly the calculations but the main problem which has to
be addressed is the presence  of matter in the bulk. The matter is
necessary to support the black string solution on the brane and
its extension into the bulk. However, there is a limit where the
matter in the bulk decouples and the general formalism developed
in this work can be applied. This issue is under investigation.

\section*{Acknowledgments}
We had many stimulating discussions with Jiro Soda, Alex Kehagias,
Christos Charmousis, Kazuya Koyama and Antonios Papazoglou. We
also thank Rodrigo Olea for correspondence. B.C.-M. is supported
by the State Scholarships Foundation (IKY) under contract 1288.

%\newpage
%\pagenumbering{Alph}
% \setcounter{page}{0}

\begin{appendix}
%\appendix{A}

\section{The Lichnerowicz equation for the scalar modes}

In this appendix we give some technical details for calculating
the scalar modes. First, we take the traceless equation, solve it
for the function $h_{rr}$, and then substitute into the
$(\theta\theta)$ component of the Lichnerowicz equation
$(\ref{ModifiedLichne})$. Also we substitute the function $h_{rr}$
into $\nabla^{A}h_{A\r}$ equation, from which we solve for
$\partial_{r}h_{r\r}$.  Then, we replace both
$\partial_{r}h_{r\r}$ and $\partial_{r}\partial_{\r}h_{r\r}$ into
the $(\theta\theta)$ equation. We follow the same steps also for
the $(\r\r)$ equation of  $(\ref{ModifiedLichne})$. From the
resulting two equations we take the combination $4 \a \b^{2}
\frac{\tanh^{2}\left(\frac{\r}{2
\sqrt{\a}}\right)}{sech^{2}\left(\frac{\r}{2 \sqrt{\a}}\right)}
(\r\r)-(\theta\theta)$ and we find the following equation for the
two scalar modes

\bea \lb{FirstScalarEq}
&&\frac{\left(2\,l^{2}\,\a\,\Omega^{2}-\left(l^{2} M-r^{2}\right)\,\cosh\left(\frac{\r}{ \sqrt{\a}}\right)\,\hbox{sech}^{4}\left(\frac{\r}{2 \sqrt{\a}}\right)\right)}{4\,l^{2}\,\left(l^{2} M-r^{2}\right)\,\a}\,h_{\theta\theta} \nn \\
&&+\b^{2}\frac{\left(3r^{2}-l^{2}\left(3M-\a\Omega^{2}\right)\right)\left(1-8\cosh\left(\frac{\r}{
\sqrt{\a}}\right)\right)-5\left(l^{2}M-r^{2}\right)\cosh\left(\frac{2\r}{
\sqrt{\a}}\right)}{8l^2\left(l^{2}\,M-r^{2}\right)\hbox{sech}^{-4}\left(\frac{\r}{2
\sqrt{\a}}\right)}\,h_{\r\r}
\nn \\
&&-\frac{6\hbox{csch}^{3}\left(\frac{\r}{
\sqrt{\a}}\right)\sinh^{2}\left(\frac{\r}{2
\sqrt{\a}}\right)}{l^{2}\,\sqrt{\a}}\,\partial_{\r}\,h_{\theta\theta}-\frac{\sqrt{\a}\b^{2}\left(-3+4\cosh\left(\frac{2\r}{
\sqrt{\a}}\right)\right)\sinh\left(\frac{\r}{2
\sqrt{\a}}\right)}{l^{2}\,\hbox{cosh}^{3}\left(\frac{\r}{2\sqrt{\a}}\right)}\,\partial_{\r}h_{\r\r}\,\,
\nn \\
&&+\frac{\partial_{\r\r}\,h_{\theta\theta}}{l^{2}\left(1+\cosh\left(\frac{\r}{\sqrt{\a}}\right)\right)}-\frac{2\a\b^{2}\tanh^{2}\left(\frac{\r}{2\sqrt{\a}}\right)\partial_{\r\r}h_{\r\r}}{l^{2}}
\nn \\
&&-\frac{\left(l^{2}M-3r^{2}\right)\hbox{sech}^{4}\left(\frac{\r}{2
\sqrt{\a}}\right)\partial_{r}h_{\theta\theta}}{2l^{4}r}-\frac{\left(l^{2}M-r^{2}\right)\hbox{sech}^{4}\left(\frac{\r}{2
\sqrt{\a}}\right)\partial_{rr}h_{\theta\theta}}{2l^{4}}\nn\\
&&+\frac{32\left(l^{2}M-3r^{2}\right)\a\b^{2}\hbox{csch}^{4}\left(\frac{\r}{\sqrt{\a}}\right)\sinh^{6}\left(\frac{\r}{2\sqrt{\a}}\right)\partial_{r}h_{\r\r}}{l^{4}r}
\nn\\
&&+\frac{32\left(l^{2}M-r^{2}\right)\a\b^{2}\hbox{csch}^{4}\left(\frac{\r}{\sqrt{\a}}\right)\sinh^{6}\left(\frac{\r}{2\sqrt{\a}}\right)\partial_{rr}h_{\r\r}}{l^{4}}=0\,\,
\eea

We need one more equation in order to find  a solution for the
scalar perturbations. The second equation comes from the $(t\r)$
component of  $(\ref{ModifiedLichne})$, and it can be derived in a
similar fashion as (\ref{FirstScalarEq}). Again we solve the
traceless condition for $h_{rr}$ and substitute into the
$(t\r)$ equation. We also replace this equation  into
$\nabla^{A}h_{A\r}$ and solve for $\partial_{r}\,h_{r\r}$. We
do the same for $\nabla^{A}h_{A t}$ and solve for
$\partial_{r}h_{t\r}$, from which we get $\partial_{r}\partial_{\r}h_{t\r}$.  Finally, from the equation $(t\r)$ we get
 the following

\be \lb{SecondScalarEqa}
h_{\r\r}=\frac{1}{\b^{2}}\partial_{\r}\left(\frac{h_{\theta\theta}}{\sqrt{a}
\sinh\left(\frac{\r}{\sqrt{\a}}\right)}\right).
\ee
Using equation (\ref{SecondScalarEqa}), equation  (\ref{FirstScalarEq}) becomes

\bea \lb{Scalar1}
&&\frac{-2\a\Omega^{2}l^2+\left(8\left(r^{2}-l^{2}M\right)-6\a\Omega^{2}l^2\right)\cosh\left(\frac{\r}{\sqrt{\a}}\right)+\left(r^{2}-l^{2}M\right)\left[7+\cosh\left(\frac{2\r}{\sqrt{\a}}\right)\right]}{8l^{2}\left(l^{2}M-r^{2}\right)\a\,\sinh^{-6}\left(\frac{\r}{2\sqrt{\a}}\right)}\,h_{\theta\theta}
\nn \\
&&\frac{-8\a\Omega^{2}l^2+\left(6\left(r^{2}-l^{2}M\right)+4\a\Omega^{2}l^2\right)\cosh\left(\frac{\r}{\sqrt{\a}}\right)+\left(r^{2}-l^{2}M\right)\left[11-5\cosh\left(\frac{2\r}{\sqrt{\a}}\right)\right]}{16\,l^{2}\left(l^{2}M-r^{2}\right)\sqrt{\a}\,\hbox{csch}^{-1}\left(\frac{\r}{2\sqrt{\a}}\right)\sinh^{-5}\left(\frac{\r}{2\sqrt{\a}}\right)}\,\partial_{\r}h_{\theta\theta}
\nn \\
&&-\frac{\left[-7+\cosh\left(\frac{\r}{\sqrt{\a}}\right)\right]\hbox{sech}^{4}\left(\frac{\r}{2\sqrt{\a}}\right)}{4l^{2}}\,\partial_{\r\r}h_{\theta\theta}
%\nn \\
%&&
-\frac{2\sqrt{\a}\,\hbox{csch}\left(\frac{\r}{\sqrt{\a}}\right)\tanh^{2}\left(\frac{\r}{2\sqrt{\a}}\right)}{l^{2}}\,\partial_{\r\r\r}h_{\theta\theta}\,
\nn \\
&&-\frac{\left(l^{2}M-3r^{2}\right)\left[1+3\,\cosh\left(\frac{\r}{\sqrt{\a}}\right)\right]\hbox{sech}^{6}\left(\frac{\r}{2\sqrt{\a}}\right)}{4l^{4}r}\,\partial_{r}h_{\theta\theta}\,
\nn \\
&&+\frac{\left(l^{2}M-3r^{2}\right)\,\sqrt{\a}\,\hbox{sech}^{4}\left(\frac{\r}{2\sqrt{\a}}\right)\,\tanh\left(\frac{\r}{2\sqrt{\a}}\right)}{l^{4}\,r}\,\partial_{r}\,\partial_{\r}h_{\theta\theta}\,
\nn \\
&&-\frac{\left(l^{2}M-r^{2}\right)\left[1+3\,\cosh\left(\frac{\r}{\sqrt{\a}}\right)\right]\hbox{sech}^{6}\left(\frac{\r}{2\sqrt{\a}}\right)}{4\,l^{4}}\,\partial_{rr}h_{\theta\theta}\,
\nn \\
&&+\frac{\left(l^{2}M-r^{2}\right)\,\sqrt{\a}\,\hbox{sech}^{4}\left(\frac{\r}{2\sqrt{\a}}\right)\,\tanh\left(\frac{\r}{2\sqrt{\a}}\right)}{l^{4}}\,\partial_{rr}\,\partial_{\r}\,h_{\theta\theta}=0\,\,
\eea
Inspecting equation (\ref{Scalar1}) we can see that it has a
third derivative with respect to $\r$ plus  mixed derivatives
$\partial_{r}\partial_{\r}$ and $\partial_{rr}\partial_{\r}$.
 This makes the handling of this equation extremely difficult
 even for numerical methods. To proceed we take the following
 Ansatz for the function $h_{\theta\theta}$
 \be \lb{hthetathetaa}
h_{\theta\theta}=\left[2\,\b\,\sqrt{\a}\,\sinh\left(\frac{\r}{2
\sqrt{\a}}\right)\right]^{2} u\left(r,\r\right).
\ee
Then, by substituting (\ref{hthetathetaa}) into (\ref{Scalar1}) we obtain

\bea \lb{Scalar2}
&&\frac{2\,\a\,\Omega^{2}\,\hbox{sech}^{2}\left(\frac{\r}{2\,\sqrt{\a}}\right)\,\tanh^{4}\left(\frac{\r}{2\,\sqrt{\a}}\right)}{l^{2}M-r^{2}}\,u\left(r,\r\right)
\nn \\
&&-\frac{\left[6\left(r^{2}-l^{2}M\right)+8\a\Omega^{2}l^2\right]\left[1+\cosh\left(\frac{\r}{\,\sqrt{\a}}\right)\right]-3\left(r^{2}-l^{2}M\right)\left[1+\cosh\left(\frac{2\r}{\,\sqrt{\a}}\right)\right]}{4l^2\a^{-\frac{1}{2}}\,\hbox{sech}^{-4}\left(\frac{\r}{2\,\sqrt{\a}}\right)\,\tanh^{-1}\left(\frac{\r}{2\,\sqrt{\a}}\right)\left(l^{2}M-r^{2}\right)}\,\partial_{\r}u\left(r,\r\right)
\nn \\
&&-\frac{\a\left[-1+7\cosh\left(\frac{\r}{\,\sqrt{\a}}\right)\right]\hbox{sech}^{2}\left(\frac{\r}{2\,\sqrt{\a}}\right)\,\tanh^{2}\left(\frac{\r}{2\,\sqrt{\a}}\right)}{l^{2}}\,\partial_{\r\r}u\left(r,\r\right)\,\,
\nn \\
&&-\frac{4\,\a^{\frac{3}{2}}\tanh^{3}\left(\frac{\r}{2\,\sqrt{\a}}\right)}{l^{2}}\,\partial_{\r\r\r}u\left(r,\r\right)\,\,
\nn \\
&&-\frac{2\,\left(l^{2}M-3r^{2}\right)\,\a\,\hbox{sech}^{2}\left(\frac{\r}{2\,\sqrt{\a}}\right)\,\tanh^{4}\left(\frac{\r}{2\,\sqrt{\a}}\right)}{l^{4}\,r}\,\partial_{r}
u\left(r,\r\right)\,\,
\nn \\
&&-\frac{2\,\left(l^{2}M-3r^{2}\right)\,\a\,\hbox{sech}^{2}\left(\frac{\r}{2\,\sqrt{\a}}\right)\,\tanh^{4}\left(\frac{\r}{2\,\sqrt{\a}}\right)}{l^{4}}\,\partial_{rr}
u\left(r,\r\right)\,\,
\nn \\
&&+\frac{128\,\left(l^{2}M-3r^{2}\right)\,\a^{\frac{3}{2}}\,\hbox{csch}^{5}\left(\frac{\r}{\,\sqrt{\a}}\right)\,\sinh^{8}\left(\frac{\r}{2\,\sqrt{\a}}\right)}{l^{4}\,r}\,\partial_{r\r}
u\left(r,\r\right)\,\,
\nn \\
&&+\frac{128\,\left(l^{2}M-3r^{2}\right)\,\a^{\frac{3}{2}}\,\hbox{csch}^{5}\left(\frac{\r}{\,\sqrt{\a}}\right)\,\sinh^{8}\left(\frac{\r}{2\,\sqrt{\a}}\right)}{l^{4}}\,\partial_{rr\r}
u\left(r,\r\right)=0\,\,
\eea
After some algebra the above equation takes  the following elegant form,
\bea \lb{Scalar3}
&&-
%\hbox{sech}\left(\frac{\r}{2\,\sqrt{\a}}\right)\,\tanh^{3}\left(\frac{\r}{2\,\sqrt{\a}}\right)
\,\frac{\partial\,}{\partial\r}
\left\{
\frac{2\,\Omega^{2}\,l^2}{\left(l^{2}M-r^{2}\right)\cosh\left(\frac{\r}{2\,\sqrt{\a}}\right)}\,u\left(r,\r\right)\right.
%\nn \\
%&&+
%\hbox{sech}\left(\frac{\r}{2\,\sqrt{\a}}\right)\,\tanh^{3}\left(\frac{\r}{2\,\sqrt{\a}}\right)
%\frac{\partial}{\partial\,\r}\left(
+\frac{3\,\left[1-2\,\hbox{cosh}^{2}\left(\frac{\r}{2\,\sqrt{\a}}\right)\right]}{\sqrt{\a}\,\sinh\left(\frac{\r}{2\,\sqrt{\a}}\right)}\,\frac{\partial}{\partial\,\r}\,u\left(r,\r\right)
%\right)
\nn \\
&&-
%\hbox{sech}\left(\frac{\r}{2\,\sqrt{\a}}\right)\,\tanh^{3}\left(\frac{\r}{2\,\sqrt{\a}}\right)
%\frac{\partial}{\partial\,\r}\left(
2\,\cosh\left(\frac{\r}{2\,\sqrt{\a}}\right)\,\frac{\partial^{2}\,u\left(r,\r\right)}{\partial
\r^{2}}
%\right)
%\nn \\
%&&+
%\hbox{sech}\left(\frac{\r}{2\,\sqrt{\a}}\right)\,\tanh^{3}\left(\frac{\r}{2\,\sqrt{\a}}\right)
%\frac{\partial}{\partial\r}\left(
\left. +\frac{2\,\frac{\partial}{\partial\,r}\left[r\,\left(r^{2}M-r^{2}\right)\frac{\partial\,u\left(r,\r\right)}{\partial\,r}\right]}{l^{2}\,r\,\cosh\left(\frac{\r}{2\,\sqrt{\a}}\right)}
%\right)
\right\} =0\,\, \eea This equation is fully integrable in the $\r$
coordinate. By setting the integration constant to zero and
separating the variables with the Ansatz
$u\left(r,\r\right)=f\left(r\right)\,y\left(\r\right)$, we get the
following two  differential equations,

\be \lb{KKmodesEq} \frac{4
\,\a^{3/2\,}\coth\left(\frac{\r}{2\,\sqrt{\a}}\right)}{l^{2}}\,\left(3\,\cosh\left(\frac{\r}{\sqrt{\a}}\right)\,\frac{\partial\,y\left(\r\right)}{\partial\,\r}+\sqrt{\a}\,\sinh\left(\frac{\r}{\,\sqrt{\a}}\right)\,\frac{\partial^{2}\,y\left(\r\right)}{\partial\,\r^{2}}
\right)+m\,y\left(\r\right)=0,
\ee
with $m$ being the variable separation constant.
%(the mass of the KK modes and}

\be \lb{FinalScalar1Eq}
\left(m-\frac{8\,\a^{2}\,\Omega^{2}}{l^{2}\,M-r^{2}}\right)\,f\left(r\right)+\frac{8\,
\a^{2}}{l^{4}\,r}\left(\left(l^{2}\,M-3r^{2}\right)\frac{\partial\,f\left(r\right)}
{\partial\,r}+r\left(l^{2}\,M-r^{2}\right)\,\frac{\partial^{2}\,f\left(r\right)}{\partial\,r^{2}}\right)=0.
\ee

\end{appendix}

\end{document}